\documentclass[twocolumn]{aastex631}
\usepackage{amsthm}
\usepackage{amsmath}
\usepackage{blindtext}
\usepackage{xcolor}
\usepackage{float}
\usepackage{graphicx}
\graphicspath{{./}{figures/}}
\begin{document}

\title{Magnification bias reveals severe contamination in Hubble Frontier Field photo-z catalogs }

\shorttitle{}
\shortauthors{Zhang et al.}

\correspondingauthor{Jiashuo Zhang}
\email{jszhang@hku.hk, joshua.z.0211@gmail.com}

\author[0000-0002-3783-4629]{Jiashuo Zhang}
\affiliation{University of Hong Kong, Hong Kong S.A.R., China}

\author[0000-0003-4220-2404]{Jeremy Lim}
\affiliation{University of Hong Kong, Hong Kong S.A.R., China}

\author[0000-0002-8785-8979]{Tom Broadhurst}
\affiliation{University of the Basque Country, Bilbao, Spain}
\affiliation{Donostia International Physics Center (DIPC), Donostia, Spain}
\affiliation{Ikerbasque, Basque Foundation for Science, Bilbao, Spain }

\author[0000-0002-4490-7304]{Sung Kei Li} 
\affiliation{University of Hong Kong, Hong Kong SAR, China}

\author{Man Cheung Li}
\affiliation{University of Hong Kong, Hong Kong SAR, China}

\author[0000-0001-8220-2324]{Giorgio Manzoni}
\affiliation{Jockey Club Institute for Advanced Study, The Hong Kong University \\
of Science and Technology, Hong Kong S.A.R., China}

\author[0000-0001-8156-6281]{Rogier Windhorst}
\affiliation{School of Earth and Space Exploration, Arizona State University, Tempe, AZ, USA.}

%% Note that the \and command from previous versions of AASTeX is now
%% depreciated in this version as it is no longer necessary. AASTeX 
%% automatically takes care of all commas and "and"s between authors names.

%% AASTeX 6.31 has the new \collaboration and \nocollaboration commands to
%% provide the collaboration status of a group of authors. These commands 
%% can be used either before or after the list of corresponding authors. The
%% argument for \collaboration is the collaboration identifier. Authors are
%% encouraged to surround collaboration identifiers with ()s. The 
%% \nocollaboration command takes no argument and exists to indicate that
%% the nearby authors are not part of surrounding collaborations.

%% Mark off the abstract in the ``abstract'' environment. 
\begin{abstract}

Gravitational lensing by massive galaxy clusters enables faint distant galaxies to be more abundantly detected than in blank fields, thereby allowing one to construct galaxy luminosity functions (LFs) to an unprecedented depth at high redshifts. Intriguingly, photometric redshift catalogs (e.g. \cite{Shipley2018}) constructed from the Hubble Frontier Fields survey display an excess of z$\gtrsim$4 galaxies in the cluster lensing fields and are not seen in accompanying blank parallel fields. The observed excess, while maybe a gift of gravitational lensing, could also be from misidentified low-z contaminants having similar spectral energy distributions as high-z galaxies. In the latter case, the contaminants may result in nonphysical turn-ups in UV LFs and/or wash out faint end turnovers predicted by contender cosmological models to $\Lambda$CDM. Here, we employ the concept of magnification bias to perform the first statistical estimation of contamination levels in HFF lensing field photometric redshift catalogs. To our great worry, while we were able to reproduce a lower-z lensed sample, it was found $\sim56\%$ of $3.5 < z_{phot} < 5.5$ samples are likely low-z contaminants! Widely adopted Lyman Break Galaxy-like selection rules in literature may give a 'cleaner' sample magnification bias-wise but we warn readers the resulting sample would also be less complete. Individual mitigation of the contaminants is arguably the best way for the investigation of faint high-z Universe, and this may be made possible with JWST observations.

\end{abstract}

%% Keywords should appear after the \end{abstract} command. 
%% The AAS Journals now uses Unified Astronomy Thesaurus concepts:
%% https://astrothesaurus.org
%% You will be asked to selected these concepts during the submission process
%% but this old "keyword" functionality is maintained in case authors want
%% to include these concepts in their preprints.

\keywords{Magnification Bias, Lensing Field UV LF, Misidentification Issue}

%% From the front matter, we move on to the body of the paper.
%% Sections are demarcated by \section and \subsection, respectively.
%% Observe the use of the LaTeX \label
%% command after the \subsection to give a symbolic KEY to the
%% subsection for cross-referencing in a \ref command.
%% You can use LaTeX's \ref and \label commands to keep track of
%% cross-references to sections, equations, tables, and figures.
%% That way, if you change the order of any elements, LaTeX will
%% automatically renumber them.
%%
%% We recommend that authors also use the natbib \citep
%% and \citet commands to identify citations.  The citations are
%% tied to the reference list via symbolic KEYs. The KEY corresponds
%% to the KEY in the \bibitem in the reference list below. 

%%%%%%%%%%%%%%%%%%%%%%%%%%%%%%

\section{Introduction} 
\label{Intro_section}
%\section{Introduction} 

%\textcolor{red}{WHY do we need cluster fields : for luminosity functions and high z searches}

%To probe the galaxy luminosity function at lower luminosities than would otherwise be possible, thus better accessing the larger population of galaxies perhaps primarily responsible for reionizing the universe \citep{Bunker2004, Robertson2013} as well as to better leverage tests for different forms of Dark Matter \citep{Corasaniti2017, Leung2018}, .  

%\textcolor{red}{Problems in using cluster fields, the most severe being misidentification}

The galaxy luminosity function (LF) - the number density of galaxies as a function of luminosity - is one of the most fundamental observables for testing cosmological models of structure formation and growth. Low-luminosity galaxies at high redshifts are of particular interest for assessing potential drivers of reionization \citep{Bunker2004, Robertson2013} as well as for leveraging tests for different types of dark matter \citep{Corasaniti2017, Leung2018}. To better detect low-luminosity galaxies at high redshifts, observations have increasingly turned to galaxy clusters for use as gravitational lenses. For example, the search for high-redshift galaxies was a primary goal of the Cluster Lensing And Supernova Survey with Hubble (CLASH) program,  comprising imaging of twenty-five galaxy clusters with the \textit{Hubble Space Telescope} (HST) in sixteen filters spanning near-UV to near-IR. The CLASH program motivated other similar programs with the HST. This includes Hubble Frontier Fields (HFF) program providing very deep images of six galaxy clusters, and Reionization Lensing Cluster Survey (RELICS) program providing shallow images of 41 clusters. More recently from imaging at near- to mid-infrared wavelengths, \textit{James Webb Space Telescope} (JWST) have showcased the ability of gravitational lensing to unveil galaxies in the early universe, see for example the work of \cite{Bradley2023,Atek2023,Atek2023B}. 

\begin{figure}
\centering	
\includegraphics[width=0.46\textwidth]{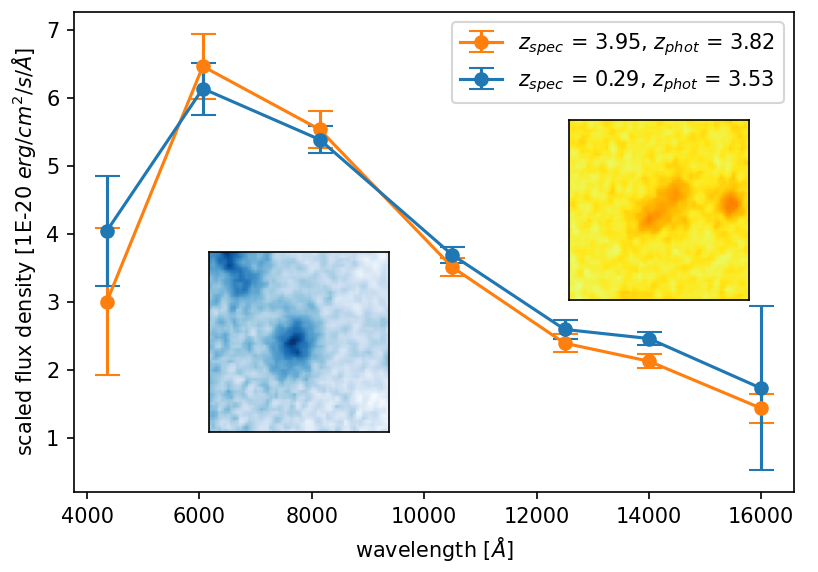}
\caption{Comparison of SED in Hubble Frontier Fields filters of a $z_{spec}$=0.29 galaxy (blue curve) with a $z_{spec}$=3.95 galaxy (orange curve). The values are taken from S18 catalog and the SED for low spec-z source is scaled up by a factor of 2.75 for a better comparison of shapes. The scaled SEDs are seen to be very similar in shape, which leads to misidentification problem of the low-z sources as is demonstrated here by their photo-z estimates. In insert figures, we show also a 1.8 arcsec$^2$ F160W band cutout image (using similar color as their SEDs) for both galaxies. Both galaxies are of similar apparent size, but the high-z galaxies are seen to be slightly more elongated. This slight visual difference, however, is not enough to refute the wrongly estimated photo-z for the low spec-z galaxy : the blue cutout could also be interpreted as the lensed image of a compact high-z galaxy affected less by lensing shear. }
\label{clusz_vs_highz_SED}
\end{figure}

While there are evident benefits in exploiting the strong lensing power of galaxy clusters, there are also inherent disadvantages associated with searches for high-redshift galaxies behind these clusters. First, gravitational lensing magnifies not only objects but also the surrounding sky, leading to a reduced observable area. In addition, the presence of abundant bright foreground cluster members can obscure dimmer background galaxies, further reducing the effective search area. Finally and perhaps most critically, without the benefit of spectroscopy, dim cluster members can be easily mistaken for high-redshift galaxies. Specifically, over the wavelength range spanned by the HST, the redshifted Balmer/4000Å breaks of quiescent galaxies at low redshift can resemble the redshifted Lyman break of star-forming galaxies at high redshifts. This situation is depicted in Figure \ref{clusz_vs_highz_SED} for a cluster member galaxy at z = 0.29 and star-forming galaxy at z = 3.95. Although both the CLASH and HFF fields benefit from observations with the Multi-Unit Spectroscopic Explorer (MUSE) at the Very Large Telescope (VLT), these observations are not sufficiently deep for measuring the redshifts of relatively dim galaxies. The lack of sufficiently deep spectroscopy is particularly pronounced for the HFF program, which provides the deepest images towards galaxy clusters available from the HST.

%The possibility that galaxy catalogs constructed for the HFF program – having redshifts inferred from SEDs, with few high-z candidates confirmed through spectroscopy – likely to suffer a significant degree of contamination have been raised by \cite{Leung2018}. 

A number of catalogs have been compiled for the HFF program, in which redshifts are inferred from SEDs with few high-z candidates confirmed by spectroscopy. The possibility that such HFF galaxy catalogs are likely to suffer contamination has been raised by \cite{Leung2018}. In particular, they noticed a significant overlap in the color-color and color-magnitude diagrams between the cluster member and the $z\sim 4$ candidates cataloged by \cite{Coe2015} (C15), suggesting a potential cross-contamination. Finding a lesser degree of overlap with cluster member above $z\leq 4.75$, Leung et al 'mitigated' this potential contamination issue in their analysis by focusing only on the $z\geq 4.75$ sample.

\begin{figure*}
\centering	
\includegraphics[width=0.97\textwidth]{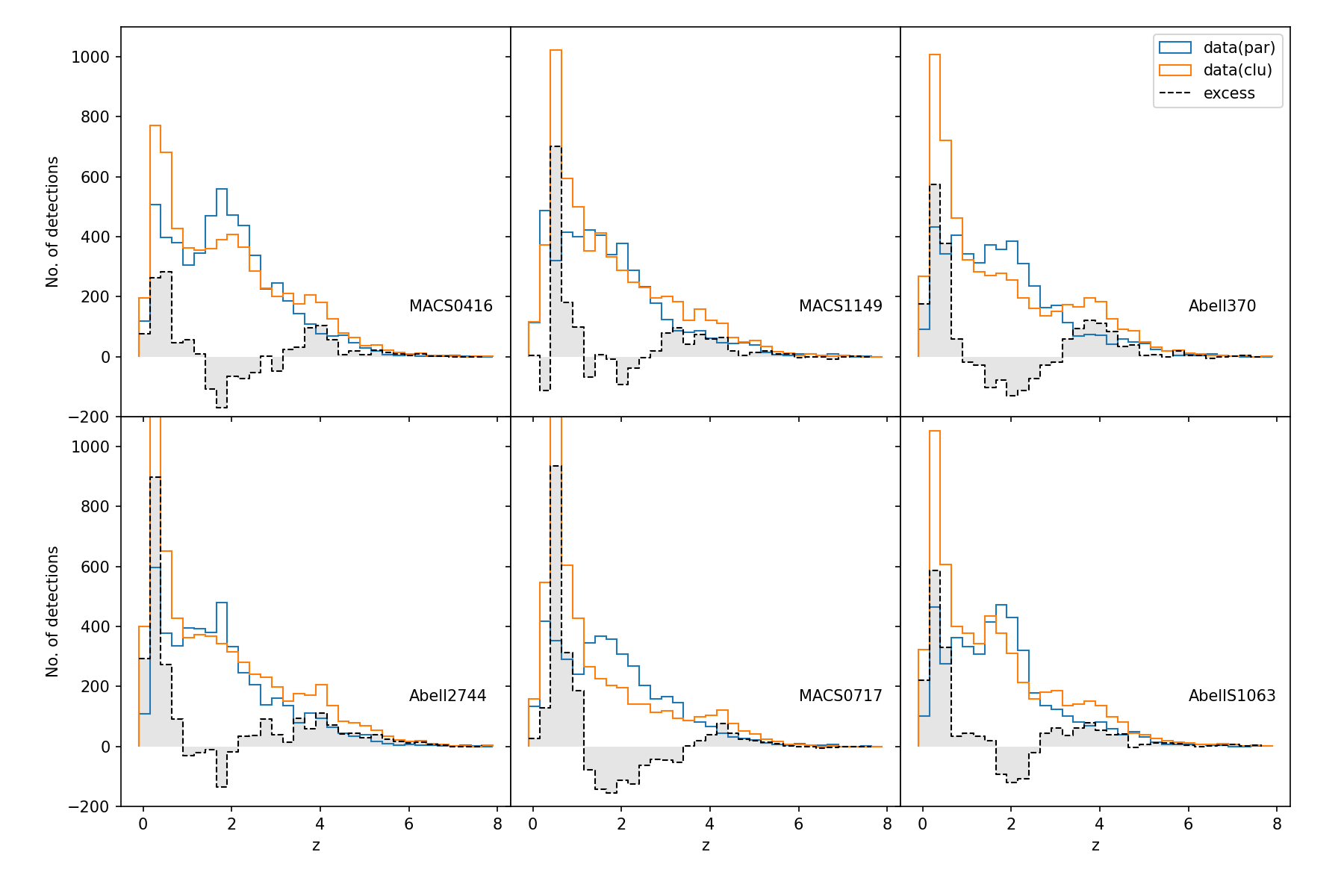}
\caption{Histogram of photometric redshifts in HFF cluster fields (orange) compared with respective parallel fields (blue) from S18 catalog. Following \cite{Shipley2018}, here we have only used sources that are non-stellar objects, with reasonable photometric redshift fitting, i.e. sources flagged with \texttt{use$\_$phot}=1. For a better visual comparison, we plot also the difference between histograms as shaded black steps. The shaded black regions for each pair of fields could be seen as composed of three parts. A part where we have excess in the cluster field at the cluster redshift, this comes from cluster members. A part with a deficit in the cluster around redshift $\sim2$, which we will show is the result of negative magnification bias over this redshift range in Sec.\ref{bias}. Finally, a second excess part above redshift $\sim$3.5, which we will argue in this paper as likely coming from the misidentification problem of low-z passive galaxies. We will also provide a quantitative estimate on the contamination fraction in Sec.\ref{bias}. }
% ---- chatgpt version
% For the sake of more uniform detection threshold, we restrict to sources outside the exclusion region defined in Sec.\ref{region_masking}. What we observe for each cluster is two concentrated cluster field bumps and a region of deficit. The first bump is at low redshift and corresponds to cluster members that remained after applying exclusion region. As we will show, the deficit in cluster fields around redshift$\sim$2 is the result of gravitational lensing decreasing surface density of detectable galaxies. At redshift $\sim4$, the bump could likewise be caused by gravitational lensing or by misidentification of cluster members. With this paper, we will argue it is actually the later. To better relate $z\sim4$ excess with lensing strength, we have adopted and marked in figure the lensing power indicator as provided by \cite{Lotz2017}: probability of a $z=9.6$ galaxy lensed to be detectable in F160W band. It is seen a stronger lens is not necessarily associated with a stronger bump, though generally could be associated with a stronger deficit. }
\label{excess}
\end{figure*}

More recently, \cite{Shipley2018} (S18) reported an excess in the number of $z\sim 4$ galaxies cataloged in HFF cluster fields than in accompanying parallel fields - nearby blank regions located 6 arcmin away from each cluster field. The reported excess is reproduced in Fig.\ref{excess}, which shows the number of galaxies\footnote{Here, and through out the paper, we use only those flagged with \texttt{use$\_$phot}=1, which are non-star-like galaxies considered to have reasonable photometries and photo-z estimations.} against redshift for each cluster (orange histogram) and parallel field (blue histogram). \cite{Shipley2018} attribute this enhancement to intrinsically low-luminosity galaxies at high redshifts made detectable by the magnification power of gravitational lensing, and claim that similar enhancements are seen in the ASTRODEEP catalogs compiled by \cite{Merlin2016}, \cite{Castellano2016}, \cite{DiCriscienzo2017} for the HFF fields.

%\textcolor{red}{What we will do to test for misidentification, and key result}

%    In \cite{Leung2018}, magnification bias was applied for a different purpose - testing wave dark matter model against the faint end behavior of UV LFs. In their work, they noticed the overlapping  

% In \cite{Leung2018}, magnification bias was applied to test wave dark matter model against the faint end behavior of UV LFs above redshift $z>4.75$. For us, however, we will use it to estimate to what extent high-redshift candidates are contaminated by low-redshift interlopers. 

In this paper, we investigate the alternative possibility of whether the excess seen in Fig.\ref{excess} is the result of severe contamination (by misidentified cluster members). To assess the plausibility of this scenario, we employ a novel test based on the concept of magnification bias. In brief here, we examine whether the number density of galaxies observed in cluster fields is compatible with the faint end extrapolation of UV LF derived from parallel fields (free of course from cluster member contamination). Using this method, we will demonstrate that about half the galaxies purported to be at $z\sim4$ are likely to be low-z interlopers. 

%we will demonstrate good compatibility at $z\sim 2$, where no contamination from cluster members is expected, but poor compatibility at $z\sim4$, suggesting significant (approximately 60\%) contamination.

This paper is organized as follows. In Sec.\ref{section_implications}, we present evidence based on various diagnostics for severe contamination among galaxies purported at $z\sim 4$. We will focus mainly on redshift range of $3.5\leq z \leq 5.5$, chosen to best capture the excess observed in the cluster fields. To more quantitatively assess the degree of contamination, we review the concept of magnification bias and describe our methodology in Sec.\ref{Method}. In Sec.\ref{bias}, we present main results of our magnification bias analysis. We would first test our methodology over redshift range of $1.2\leq z\leq 2.4$, which is expected to be free of contamination. We then extend to the redshift range of $3.5\leq z \leq 5.5$ and report the fraction of contamination among high redshift galaxies. We conclude in Sec.\ref{conclusion_section}, providing also a discussion on the impact of contaminants as well as prospects of mitigating contaminants. Throughout our analysis, we assume a simple two-component flat $\Lambda$CDM cosmology with $H_0 = 70$ km/s/Mpc and $\Omega_\Lambda = 0.7$.

%In Sec.\ref{cm_cc} we provide some diagnostics to understand the properties of those contaminants/interlopers.

\section{Comparison between cluster and parallel fields}
\label{section_implications}

%%% preambles needs to be exacting! every sentence should be exacting and deliver a sharp scientific message.

%%% ask myself whether readers can understand what is being said, context, logic etc

%%% avoid loose language

%By the end of this section, we would make statements like    the evidence is highly suggestive of the misidentified cluster members are the contaminants. but don't write it sounds like we are definitive that misidentified cluster members ARE contmainatns.

Before applying the more exacting test of magnification bias, we first assess the plausibility of the excess observed being attributed to misidentified cluster members. In Sec.\ref{sec_lensing_invariant}, we compare the color and surface brightness (chosen for their lensing-invariant nature) of $3.5\leq z\leq 5.5$ galaxy candidates in cluster fields with more robust high-z candidates in parallel fields. Any statistical discrepancy would indicate contamination. We then check if $3.5\leq z\leq 5.5$ galaxy candidates in cluster fields exhibit anticipated (1) concentration toward the cluster red sequence in Sec.\ref{sec_cm_diagram}, and (2) radial clustering in Sec.\ref{sec_radial_distribution} as misidentified cluster members.

%In this section, we assess the plausibility of severe contamination by misidentified cluster members to explain the excess observed. In Sec.\ref{sec_lensing_invariant}, we check whether $3.5\leq z\leq 5.5$ galaxy candidates in cluster fields are intrinsically different to the more robust high-z candidates in parallel fields. Any statistical discrepancy would indicate the potential contamination by low-redshift interlopers. Subsequently, in Sec.\ref{sec_cm_diagram} and Sec.\ref{sec_radial_distribution}, we examine whether $3.5\leq z\leq 5.5$ galaxy candidates in cluster fields exhibit anticipated properties as galaxy cluster members.

\subsection{color versus surface brightness}
\label{sec_lensing_invariant}

\begin{figure*}
\gridline{\fig{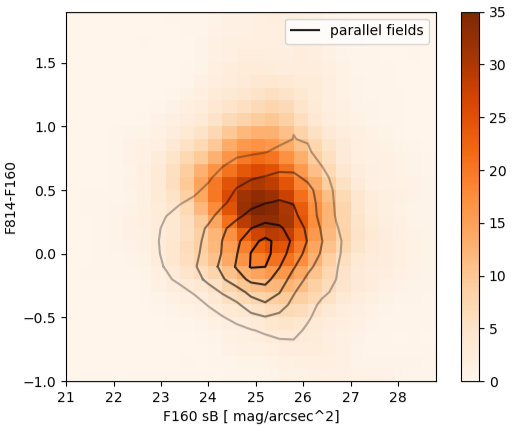}{0.435\textwidth}{(a) F814-F160 vs F160 band surface brightness.}%\label{F814160}
\fig{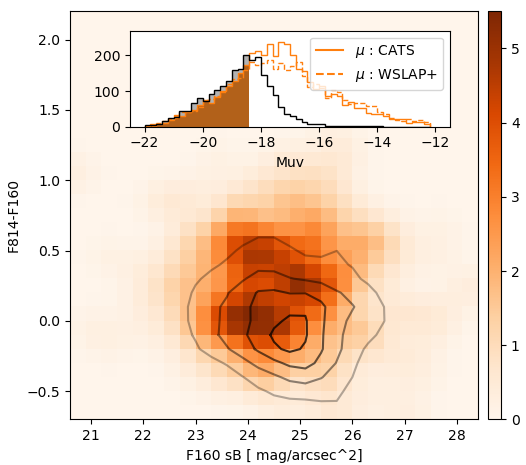}{0.435\textwidth}{(b) color vs surface brightness with $M_{uv}$ cut.}}
\caption{Comparison of intrinsic properties between $3.5\leq z_{phit}\leq5.5$ high-z candidates in the cluster and parallel fields, using only \texttt{use\_phot}=1 galaxies from S18 catalogs. Here we focus dominantly on lensing invariant observables: color F814-F160 capturing the SED redwards of the Lyman break, and F160W band surface brightness capturing their intrinsic brightness. In both figure (a) and (b), color-surface brightness distribution is shown as a background 2d histogram with a mild degree of Gaussian smoothing for the cluster fields, and as black contours of different transparency for the parallel fields. Figure (a) used all \texttt{use\_phot}=1 galaxies, and is indicating $3.5\leq z_{phit}\leq5.5$ candidates in cluster fields to be statistically redder than in parallel fields. But, as revealed by the insert of figure (b), in cluster fields we have a lot intrinsically fainter galaxies lensed to be observable: after lensing correction (dashed with WSLAP+ and solid with CATS model), the histograms of $M_{uv}$ for cluster fields (orange) peak at a lower magnitude than that for parallel fields (black). To remove those intrinsically fainter galaxies, which may also be statistically redder, we impose $M_{uv}<-18.4$ (shaded region under histograms) for figure (b). The redder population relative to parallel fields is seen persistent after the $M_{uv}$ cut, demonstrating contamination even among the UV bright sample.}
\label{color_sB_con}
\end{figure*}

\begin{figure*}
\centering
\includegraphics[width=0.97\textwidth]{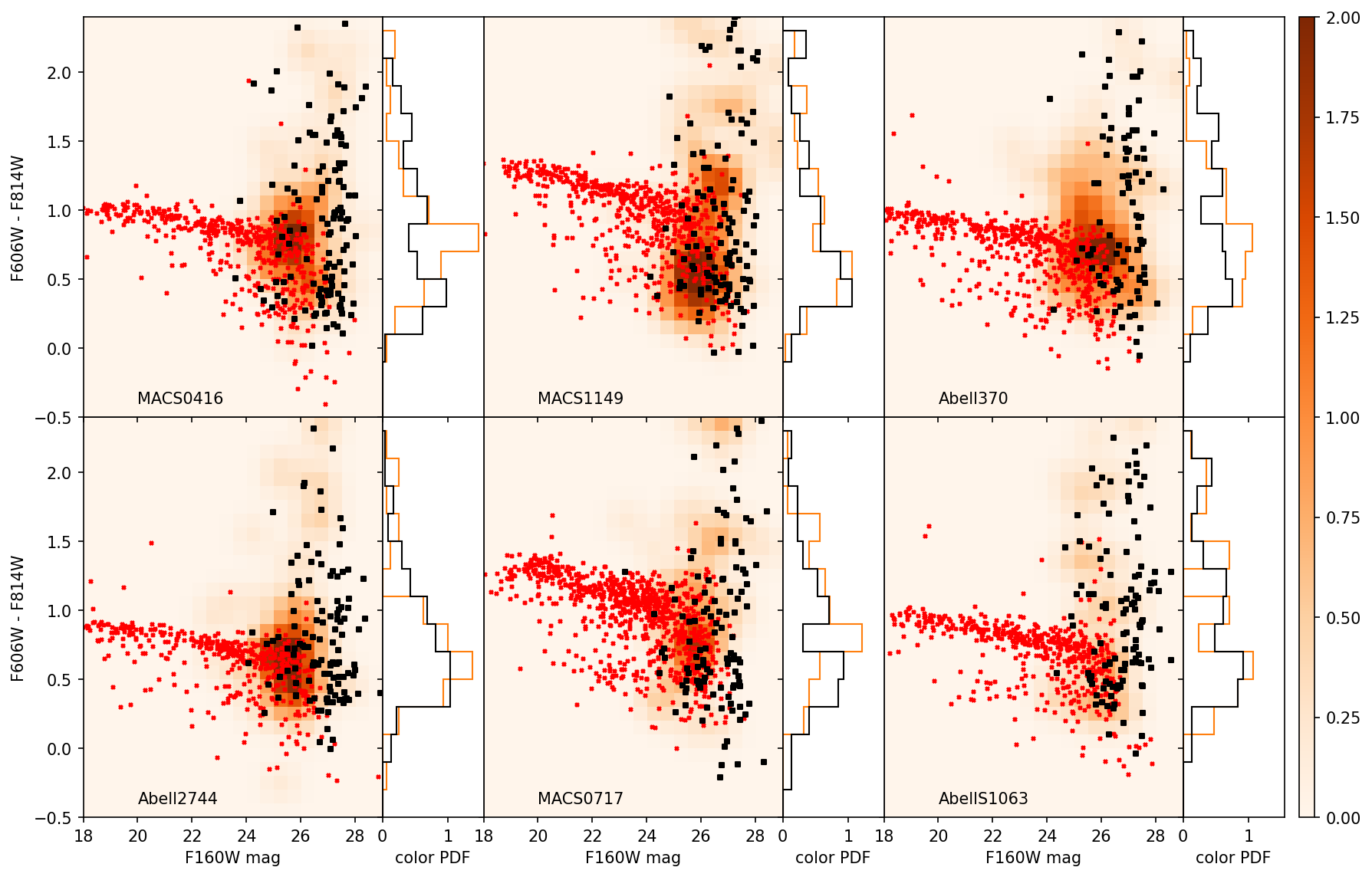}
\caption{Color-magnitude diagram for the UV-bright sample with $M_{uv}<-18.4$ to test the plausibility of misidentified cluster members as a source of contamination. Red dots are galaxies in the cluster fields with photo-z falling in $z_{clu} \pm 0.15$, and are plotted to indicate the position of cluster Red Sequence. Black points are UV bright $3.5\leq z \leq 5.5$ galaxies in the parallel fields, and the smoothed 2d histogram background marks the distribution of 3.5-5.5 galaxies in the cluster fields. To the right of each figure, we also compare the color distribution in the cluster fields (orange histogram) with that in the parallel fields (black histogram). We note there is a strong concentration in the cluster field at the faint end extension of cluster Red Sequence, demonstrating the contaminants could indeed come from misidentified cluster members. Galaxies in the parallel fields are seen to have a similar distribution in F606-F814. This simply reflects the cause of misidentification problem, that redshifted Lyman breaks of high-z galaxies fall in the same bands as Balmer breaks of the cluster members and are of similar strength. }
\label{cm_high_z}
\end{figure*}

%Galaxies included in (b) are also highlighted in the inset of (b) as the shaded region below the $M_{uv}$ histograms. The histograms in black show the parallel field distribution, while the solid orange steps indicate the cluster field distribution corrected for lensing magnification using CATS models (the dashed orange histogram obtained using WSLAP+ models shows a similar $M_{uv}$ distribution).

Color and surface brightness are two well-established lensing invariant observables, enabling a sensible comparison between the intrinsic properties of high-z galaxy candidates in lensing cluster fields and non-lensing parallel fields. In particular, we adopt F814W-F160W as color to compare SED redwards of the redshifted Lyman breaks (or Balmer breaks for interlopers). This reflects the stellar age of galaxies, and is dependent on the amount of dust extinction. For surface brightness, as a proxy for the stellar mass of galaxies, we adopt the longest wavelength band - F160W - as passive cluster members would be most easily detected. 

Fig.\ref{color_sB_con}(a) shows the corresponding color versus surface brightness diagram. The distribution of $3.5\leq z\leq 5.5$ candidates in cluster fields is shown as a background 2d histogram (after a mild degree of Gaussian smoothing), with the distribution for parallel fields shown by black contours of different transparency. As is apparent, $3.5\leq z\leq 5.5$ candidates in the cluster fields are statistically redder than those in the parallel field at similar surface brightness. Contamination by passive cluster members can explain this as they tend to be redder than high-z star-forming galaxies owing to older stellar populations. Intrinsically fainter High-z galaxies, on the other hand, may also be redder either being less star-forming or more dust-extincted. Moreover, gravitational lensing could magnify these intrinsically fainter and redder high-z galaxies to be abundantly observed in the cluster fields, providing an alternative explanation for the difference observed. 

%(e.g. in the F814W band, at a fixed F160W band surface brightness)

To remove the intrinsically fainter galaxies from consideration, we impose a cut on absolute magnitude. The insert of Fig.\ref{color_sB_con}(b) shows the distribution of absolute UV magnitudes ($M_{uv}$) at rest frame 1600$\AA$ for high-z candidates in cluster fields as orange histograms and in parallel fields as black histogram. The cluster field histograms used either CATS models (solid, see Sec.\ref{sec_mag_bias_equation} for more detail) or WSLAP+ models (dashed) to correct for lensing magnifications, and the resulting distributions are similar. As can be seen, over the selected redshift interval, galaxies in the cluster field peak at lower luminosities than those in the parallel field. The intrinsically fainter galaxies could be removed by imposing a cut at where the parallel field histogram peaks. The main panel of Fig.\ref{color_sB_con}(b) then shows the color-surface brightness comparison repeated for the remaining sample (shaded region under histograms). Evidently, the difference seen in Fig.\ref{color_sB_con}(a) between cluster and parallel fields persists, providing solid evidence for contamination even among this intrinsically bright sample. 

%the newly obtained distribution for parallel fields is similar to before the cut. But 

% whereas the distribution for cluster fields now composes of three concentrations. The bluest population is seen closest (albeit with a slight offset) to the parallel fields, suggesting they are likely genuine high-z galaxies. The offset could indicate these galaxies are statistically more massive (hence brighter) than high-z galaxies in parallel fields, and may have formed earlier (hence also slightly redder). The concentration for cluster fields seen in Fig.\ref{color_sB_con}(a) persists after the cut, with a new and even redder concentration at a brighter surface brightness. These concentrations cannot be explained with high-z galaxies, thus providing solid evidence for contamination even among this intrinsically bright sample.

\subsection{Color-magnitude diagrams}
\label{sec_cm_diagram}

%In the following sections, we further investigate whether the contamination hinted by Fig.\ref{color_sB_con}(b) could come from misidentified cluster members. For this, we note two properties of cluster members - (a) they exhibit a strong concentration towards the galaxy cluster Red Sequence in color-magnitude diagrams; and (b) they display a strong radial clustering trend towards the galaxy cluster center.

If contaminants hinted by Fig.\ref{color_sB_con}(b) are misidentified cluster members, we anticipate them to strongly concentrate towards the cluster's red sequence. Fig.\ref{cm_high_z} tests for this concentration by plotting F160W band magnitudes against the color straddling the Balmer break of cluster members - F606W-F814W. For each galaxy cluster, the red sequence and its faint end extension are indicated by $z_{clu} - 0.15 \leq z_{phot} \leq z_{clu} + 0.15$ galaxies\footnote{This range is chosen to capture the complete red sequence while avoiding features away from the red sequence.} shown as red dots. The distribution of $M_{uv}<-18.4$ $3.5 \leq z_{phot} \leq 5.5$ candidates in cluster fields are presented as the orange background, with their F606W-F814W color distributions also shown on the right as orange histograms. 

As is apparent, high-z candidates in cluster fields display prominent concentration towards the respective red sequence at the faint end, suggesting they may indeed be misidentified cluster members. High-z candidates in parallel fields shown as black data points are seen to span a similar range in color (see also the black histograms on the right). This reflects the cause of misidentification problem - that the Balmer breaks of cluster members could be easily confused as the Lyman breaks from $3.5\leq z\leq 5.5$ with a comparable strength. Neverthelesss, high-z candidates in parallel fields are seen offset from the cluster red sequence concentration as well as high-z candidates in cluster fields magnitude-wise. This further motivates the possibility of contamination in cluster fields. 

%This reflects the cause of the misidentification problem - that the redshifted Lyman breaks of high-z galaxies fall in the same bands as the Balmer breaks of cluster members and have comparable strength. 

%The identify of the third sample, however, remains unclear and it could be a mix between contaminants and intrinsically redder high-z galaxies owing to strong dust attenuation. As such, a clear distinction between contaminants and high-z sample remains unfeasible, hence we appeal to magnification bias for a more quantitative estimate on the contamination level. 

\subsection{Radial distribution}
\label{sec_radial_distribution}

\begin{figure*}
\centering
\includegraphics[width=0.97\textwidth]{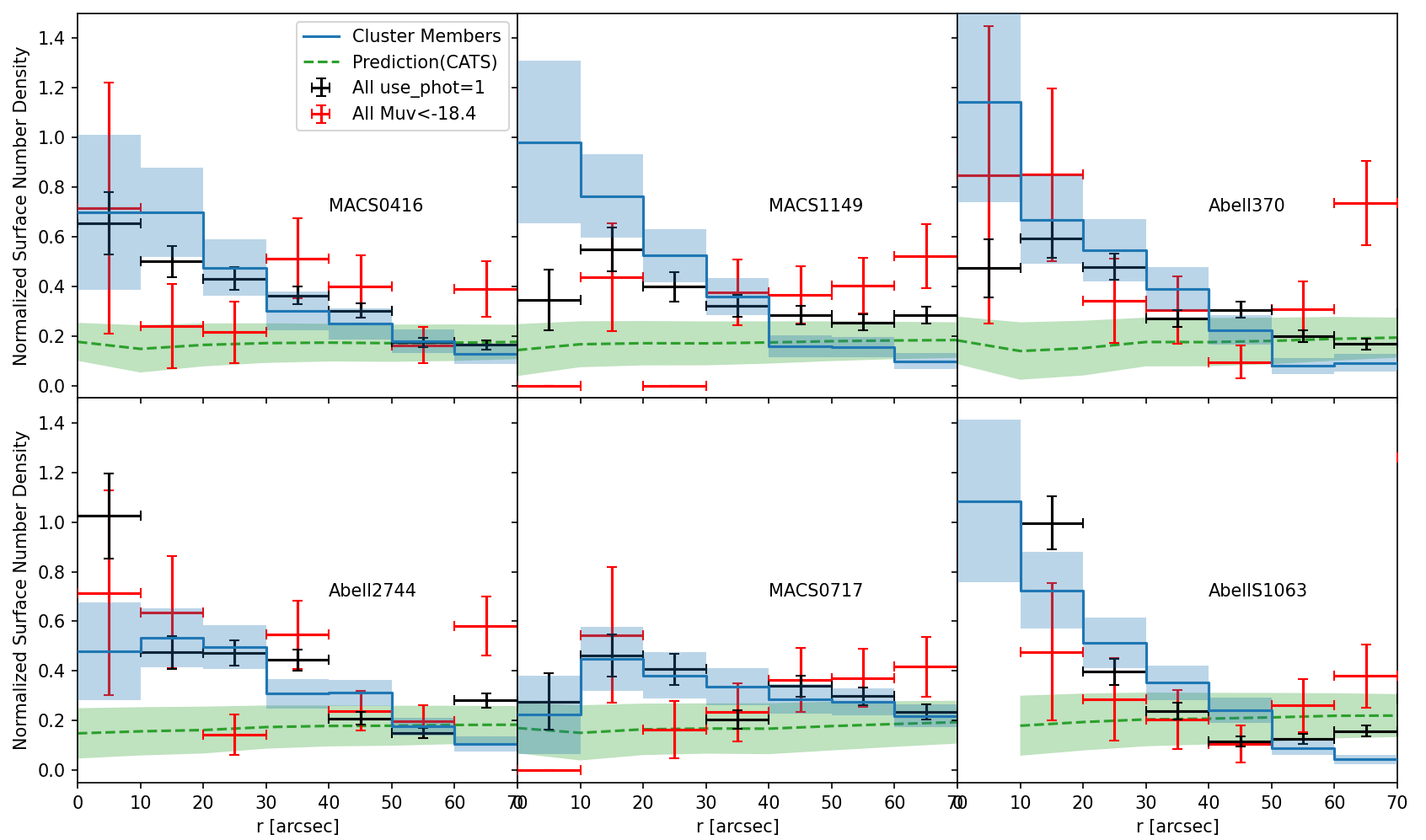}
\caption{Normalized surface Number density in different radial bins over $3.5\leq z\leq 5.5$ (black data points for all \texttt{use\_phot}=1, red data points for only UV-bright galaxies with $M_{uv}<-18.4$) compared with that for cluster members (blue step). Here by cluster members, we refer to spectroscopically confirmed galaxies with $z_{clus} - 5\sigma_z \leq z_{spec} \leq z_{clus} + 5\sigma_z$, where $\sigma_z$ is the standard deviation of the cluster redshift peak in the spectroscopic redshift histograms. $r$ is the radial distance in arc-second to the FOV center, which we adopted from \cite{Lotz2017}. The radial distributions for full \texttt{use\_phot}=1 sample are seen to follow cluster members, whereas the same clustering trend is less obvious for $M_{uv}<-18.4$ sample owing to poorer statistics. By contrast, the predicted magnification bias based on CATS model, shown as green dashed line, is seen to decrease with radius. For the $M_{uv}<-18.4$ sample, surface number density also seems to increase beyond 40 arcsec. We demonstrate in later sections that both of these distributions reflect the negative bias expected over this redshift range, suggesting $M_{uv}<-18.4$ sample to suffer less from contamination by misidentified cluster members. The later point is also consistent with our observation from Fig.\ref{color_sB_con}(b) that the contamination peak is seen as less dominant after $M_{uv}$ cut. } 
\label{cluster_member_distribution}
\end{figure*}

If contaminants are misidentified cluster members, we also anticipate them to display a radially concentrated distribution about respective cluster center. To make this check, we computed the surface number density of galaxies in 10-arcsecond wide concentric annulus bins, with the field of view (FOV) centers adopted from the \citep{Lotz2017} paper. %Note that these FOV centers may have slight offsets from the actual galaxy cluster centers. Still, the offsets are typically small enough (less than a radial bin) to be safely used for this radial clustering test.

Fig.\ref{cluster_member_distribution} plots the radial distribution (in surface number density) of all $3.5 \leq z \leq 5.5$ candidates in the cluster fields as black data points, and for $M_{uv} < -18.4$ galaxies only as red data points. For an easier visual comparison, the data points are normalized by the respective cumulative sum over different radial bins. The radial distributions for cluster members are presented as blue steps, where we define cluster members as galaxies with spectroscopic redshifts in the range $[z_{clus} - 5\sigma_z, z_{clus} + 5\sigma_z]$, with $\sigma_z$ being the standard deviation of the cluster redshift peak in the spectroscopic redshift histograms.

Notably, the measured densities using all $3.5 \leq z \leq 5.5$ candidates follow a similar radial distribution as the cluster members, adding to the evidence (Sec.\ref{sec_cm_diagram}) that most high-z candidates are likely misidentified cluster members. The same agreement with cluster members is also suggested by $M_{uv} < -18.4$ galaxies, although with a large scatter due to poorer statistics. 

By contrast, gravitationally lensed $3.5\leq z\leq 5.5$ galaxies are expected to be less populated in the inner region than in the outer region, as demonstrated by our magnification bias prediction using CATS model (green dashed line, more detail in Sec.\ref{sec_mag_bias_equation}). As we will show in Sec.\ref{bias}, this is the result of slight negative bias observed over this redshift range.

\section{Estimating Contamination Level}
\label{Method}

To quantitatively assess the degree of contamination, we will predict the surface number density of gravitationally lensed high-z galaxies and identify any excess population observed as contamination. In doing so, we make use of the concept of \textit{magnification bias}, which collectively accounts for two opposing effects by lensing magnifications on the observable surface number density of galaxies: (i) effective lowering of the detection thresholds, enabling the detection of intrinsically fainter galaxies, and (ii) the physical sky area probed given the same telescope field of view is simultaneously reduced.

In the remainder of this section, we address how we determine the detection thresholds and galactic luminosity functions needed for the above magnification bias analysis. We start in Sec.\ref{sec_data_complete} by imposing spatial uniformity in detection thresholds, a desired property for the best implementation of magnification bias. Subsequently, we determined the data completeness thresholds common to both cluster and parallel fields for the S18 catalogs. Then in \ref{sec_counts_construct}, we measure the UV LF from parallel fields, which naturally suffer less from contamination by misidentified cluster members. The detailed methodology is presented in Sec.\ref{sec_mag_bias_equation}. Finally, in Sec.\ref{subsec_uv_bright_sample}, we comment on the data-complete sample of UV-bright galaxies data (spanning a wide redshift range) needed for analysis.

%To implement magnification bias, we thus need two basic ingredients: (i) UV LF with which we can extrapolate to fainter magnitudes, and (ii) a common detection threshold for cluster and parallel field. In Sec.\ref{sec_data_complete}, we detail how we determine detection thresholds for \cite{Shipley2018}. Then in Sec.\ref{sec_counts_construct} we measure UV LFs needed from HFF parallel fields. Subsequently we detail how to apply magnification bias and predict for observable surface number densities in Sec.\ref{sec_mag_bias_equation}, and comment on the sample suitable for magnification bias analysis in Sec.\ref{subsec_uv_bright_sample}. %, given parallel fields suffer less contamination by cluster members. 

%the detection threshold defined for both cluster and parallel fields, so that we know where to start extrapolate.

\begin{figure}
\includegraphics[width=0.46\textwidth]{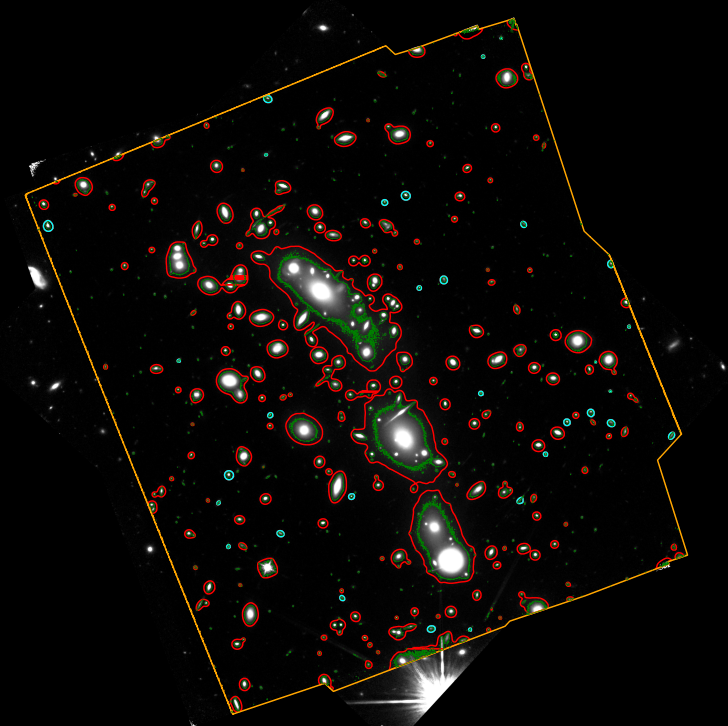}
\caption{Exclusion and common region for cluster field of M0416. To attain a more uniform detection threshold, we follow \cite{Leung2018} and mask out bright regions above $0.005$e/s on smoothed F160 images from analysis. These regions are enclosed by red contours. Cyan contours enclose bright regions above $0.005$e/s but contains a single galaxy candidate with redshift above 1.2 (as per S18 catalogs), such regions are preserved for accurately measuring galaxy number counts at the bright end. To best exploit the HFF data, we further restrict to the common HFF coverage region indicated by the orange contours. The background image is from HFF F160W band. }
\label{exclusion_region}
\end{figure}

\begin{figure*}
\centering
\includegraphics[width=0.95\textwidth]{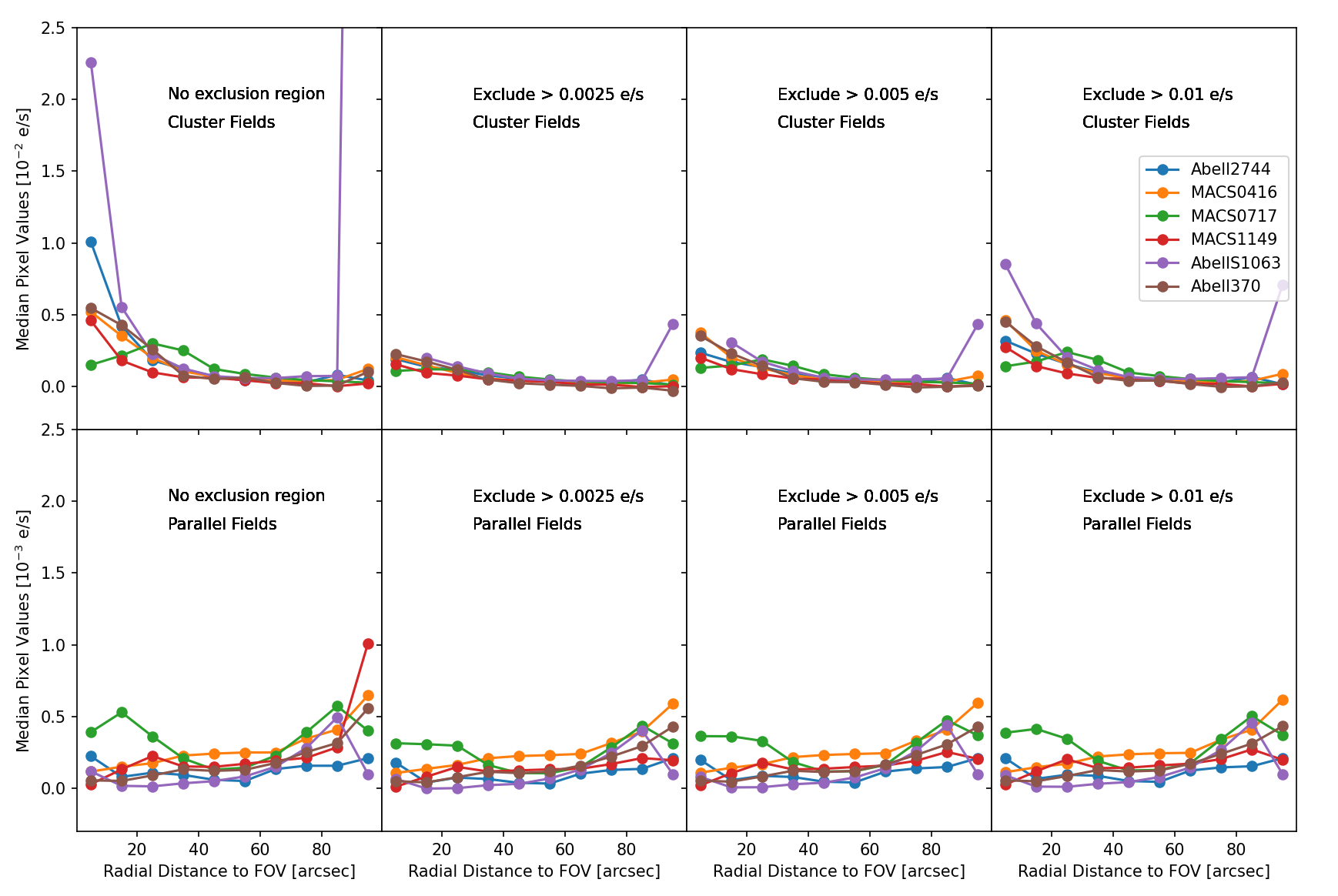}
\caption{Compare different choice to define bright regions.}
\label{compare_exclu_definition}
\end{figure*}

\begin{figure}
\includegraphics[width=0.46\textwidth]{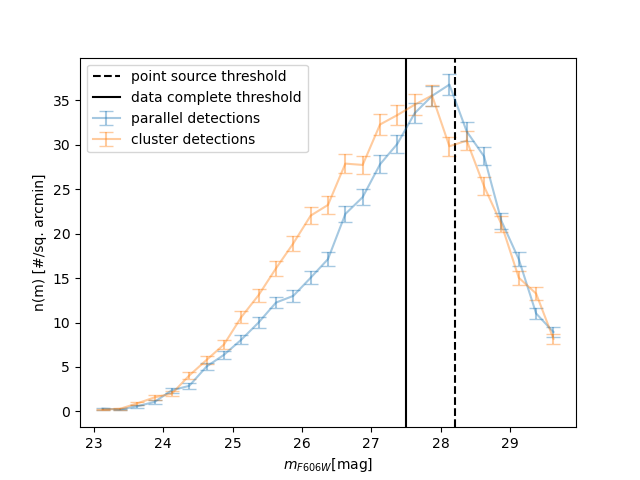}
\caption{Surface number densities of galaxies (cluster field as orange, parallel field as blue) as function of F606 band apparent magnitudes. $5\sigma$ threshold for a point source over 0.7" circular aperture is shown as black dashed line, below which parallel densities drop drastically. Cluster field densities drop at a brighter magnitude, which we attribute to galaxies of same brightness are larger than in the parallel field.  To impose the same completeness threshold in both the cluster and parallel fields, we define a conservative threshold at 0.25 mag brighter than where the cluster field number density drops dramatically, as indicated by the solid black vertical line. }
\label{neg_data_compl}
\end{figure}

\subsection{Uniform Completeness Thresholds}
\label{sec_data_complete}

In cluster fields, the observable surface number density of galaxies could be predicted by extrapolating the underlying galaxy luminosity function to magnitudes fainter than the detection threshold. The presence of bright galaxies, however, hinders the detection of dimmer galaxies and result in a non-uniform detection threshold across the cluster fields (likewise for parallel fields but to a lesser degree). To mitigate this spatial variation, we mask out bright regions in both cluster and parallel fields as described below. 

Fig.\ref{exclusion_region} shows one of the cluster field - MACSJ0416 - as imaged by the HFF F160W band\footnote{HFF (v1 release, 30mas) data used in this paper can be found in MAST: \dataset[10.17909/T9KK5N]{http://dx.doi.org/10.17909/T9KK5N}. WSLAP+ and CATS models adopted in this paper could also be found on the lens model page.}. In this band, bright cluster cluster members are most easily detected, allowing for a best identification of the bright regions. In Fig.\ref{exclusion_region}, we identify bright regions above pixel unit of 0.01$e/s$  (corresponding to brightness threshold of 23.33 mag/arcsec$^2$) as regions enclosed by green contours. These contours are seen to mostly enclose bright cluster members, and display fine-scale structure tracing non-physical noise peaks around bright objects. To smooth over the noise peaks, we apply smoothing to F160W band images with a Gaussian kernel of size (standard deviation) 16 pixels. Examples of smoother contours are shown in red and cyan, for which we used a different brightness threshold (0.005$e/s$, or 24.08 mag/arcsec$^2$) for a clearer visual comparison. Cyan contours enclose bright regions containing a single $z_{phot} > 1$ galaxy, such regions will be preserved for analysis.  

%Subsequently to find the optimal brightness cut on Gaussian smoothed images, we partitioned each field into ten concentric radial bins and plotted the median pixel values in remaining regions against radial distance to the center of field of view.
To reflect upon a more uniform detection threshold, we test if the median brightness level is similar across the remaining region. 
Fig.\ref{compare_exclu_definition} shows the median brightness level measured in different radial annulus bins for individual cluster and parallel fields. Here we compare the resulting distribution for three different brightness thresholds as well as for the case without any brightness threshold. As could be seen, the measured distributions in parallel fields are all relatively flat in all four panels. In the cluster fields, the most stringent cut of 0.0025$e/s$ (24.83 mag/arcsec$^2$) results in the flattest distribution. Motivated by this flatness, the choice of 0.0025$e/s$ was adopted by \cite{Leung2018} to remove most of the galaxy cluster lights for their UV LF faint end test above redshift $z>4$. For us, however, we are interested in the misidentified cluster members which are more populated towards the inner region. Hence to preserve a larger sample and better test for contamination, we choose a less stringent cut of 0.005$e/s$. In Fig.\ref{exclusion_region}, the final exclusion regions thus defined for M0416 are shown as regions enclosed by red contours. Note with the choice of 0.005$e/s$, there remains a noticeable increase in brightness level for each cluster fields in the innermost region. The implication of this rise will be discussed in result sections. 

To exploit the full availability of HFF photometric measurements, we further restrict our analysis to the common area covered by all HFF bands. This is shown as the region enclosed by the orange contour in Fig.\ref{exclusion_region}.

After imposing exclusion regions and common regions, we determine data completeness threshold for each band by plotting measured surface number density of galaxies as a function of apparent magnitudes. Fig.\ref{neg_data_compl} shows the example result for the F606W band, where the orange histogram combined all available cluster fields and blue combined all parallel fields. The number density of galaxies in the cluster fields is seen elevated to that in the parallel fields over apparent magnitudes $\sim 25-27$. This could be the result of either magnified distant galaxies, or abundant dim cluster members, or both. The 5$\sigma$ detection threshold for point sources in both fields is indicated by a dashed vertical line. As can be seen, whereas the number density of galaxies drops dramatically below this threshold in the parallel fields, the number density in the cluster fields begins to drop at about 0.3 mag brighter than this threshold. This difference is caused presumably by differences in the distribution of galaxy sizes in the two fields, such that at the same apparent magnitude there are more spatially extended galaxies in the cluster fields (e.g. extended cluster members, or lensed distant galaxies with large magnification factor) than in the parallel fields. To define the same completeness level for both fields, we conservatively adopt a magnitude of 27.5
as indicated by the vertical solid line in Fig.\ref{neg_data_compl}. Data complete thresholds in other wavelength bands are determined similarly for S18 catalogs, and are 28.0 for F435W band, 27.25 for F814W to F160W bands.  

%that is 0.25 mag (size of magnitude bin used) brighter than where the number density in the cluster field drops dramatically 

\subsection{Luminosity Function}
\label{sec_counts_construct}

\begin{figure}
\includegraphics[width=0.46\textwidth]{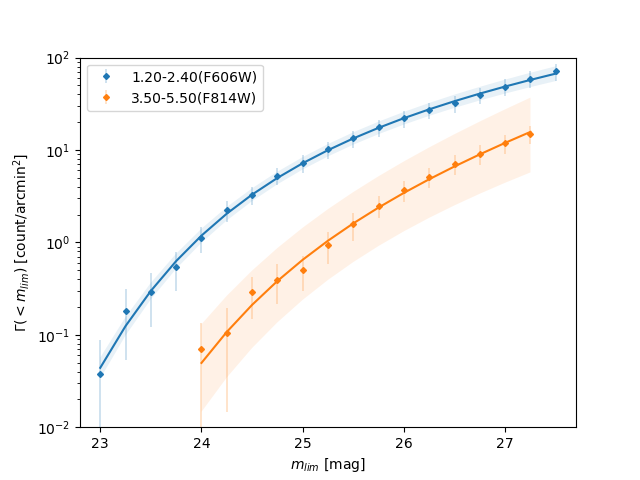}
\caption{Galaxy number counts over redshift 1.2-2.4 (blue) and 3.5-5.5 (orange). Each data point denote the unlensed density of all sources brighter than the corresponding magnitude down to data completeness threshold, where magnitudes are from the bands (indicated in legend) immediately longwards the Lyman break. Data points are averaged over measurements from six parallel fields, and error bars reflect both Poisson statistics as well as dispersion about the plotted average. Solid curves of same color are the fitted Gamma functions with shaded region corresponding to fitting uncertainties.}
\label{low_LF}
\end{figure}

\begin{figure}
\includegraphics[width=0.46\textwidth]{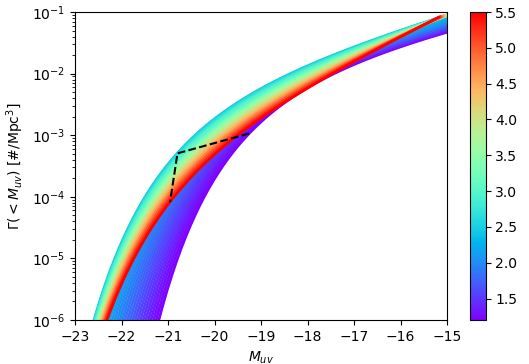}
\caption{Integrated UV LFs at different redshifts extracted from our fitted LF-redshift relations. The color bar reflects the redshift which ranges from z=1.2 to z=5.5. Black dashed line joins the position of $M_*$ at different redshift.}
\label{LF_zs}
\end{figure}

%Prior to estimating contamination fraction among $3.5 \leq z \leq 5.5$ galaxies, we wish to test magnification bias methodology over a lower redshift range - $1.2\leq z \leq 2.4$.  Over $1.2\leq z \leq 2.4$, galaxies have distinct SEDs to cluster members hence this redshift range would be less affected by misidentification problem. This redshift range is also chosen to best reflect the range over which cluster fields show a deficit in galaxies observed relative to parallel fields. To derive UV LF needed for both $3.5 \leq z \leq 5.5$ and $1.2\leq z \leq 2.4$.

In this work, we are interested in magnification bias analysis (in particular, contamination level estimates) over $3.5 \leq z \leq 5.5$. Prior to that, we first test our methodology against a lower redshift range, $1.2 \leq z \leq 2.4$, over which galaxies have distinct SEDs to cluster members hence is less likely to suffer from misidentification problem. To derive UV luminosity functions needed for both redshift ranges, in Fig.\ref{low_LF} we plot the surface number density of galaxies in the parallel fields as a function of limiting magnitudes in blue for $1.2 \leq z_{phot} \leq 2.4$ and in orange for $3.5\leq z_{phot} \leq 5.5$. Here, to cope with the convention of UV LF measurements, apparent magnitudes in the higher redshift range are measured in the band (F814W) that captures the rest frame $1600\AA$ at the average redshift of 4.5. At average redshifts $z<2$, however, redshifted Lyman breaks may fall in the same band as $1600\AA$ or be very close. Hence we adopted the band capturing rest frame 2000$\AA$ (i.e. F606W) instead for redshift range $1.2 \leq z_{phot} \leq 2.4$. 

%Plotted values in Fig.\ref{low_LF} are upon averaging over all parallel fields, with error bars reflecting both Poisson statistics as well as root-mean-square dispersion between fields (i.e., reflects cosmic variance). 

The measured functions in Fig.\ref{low_LF} are essentially the integrated form of UV LFs, and UV LFs are known to be well-parameterized by Schechter functions. Hence for magnification bias, we fit the corresponding integral form (Gamma functions $\Gamma(<m_{lim})$) to achieve extrapolation toward fainter magnitudes. Conversion to absolute magnitudes at rest frame 1600$\AA$ is done via the following equation derived from \cite{Leung2018}:

\begin{equation}
\begin{aligned}
   & M_{UV}(m_{obs}, \lambda_{eff}, z) =  m_{obs}  \\
    & + 2.5 \log_{10} \bigg{[} \frac{\mu}{(1+z)^{\beta + 1}} \bigg{(} \frac{\lambda_{eff}}{1600} \bigg{)}^{\beta + 2} \bigg{(} \frac{10 \text{pc}}{D_L(z)} \bigg{)}^{2} \bigg{]}
\end{aligned}
\end{equation}

\noindent where $\lambda_{eff}$ is the pivotal wavelength of the band used for apparent magnitudes, $D_L$ is the luminosity distance, and $\beta$ is the typical UV continuum slope of galaxy SEDs. The factor of $(\lambda_{eff}/1600)^{\beta+2}$ is needed to correctly convert absolute magnitude from rest frame $\lambda_{eff}/(1+\bar{z})$ to 1600$\AA$. For $\beta$, \cite{Bouwens2014B}, \cite{Bouwens2009}, \cite{Hathi2016} have measured its relation with $M_{UV}$ at different redshifts. By collecting these relations and fitting against redshift, we obtained $\beta(M_{UV},z)= -0.118 z -1.288 - (0.029 z + 0.003)(M_{UV}+19.5)$ for our magnitude conversion. 

In Fig.\ref{low_LF}, the best-fit Gamma functions over $1.2 \leq z \leq 2.4$ and $3.5\leq z \leq 5.5$ are plotted as solid lines, with fitting uncertainties indicated by shaded regions. Obtained Schechter parameters are also provided in the first two rows of Table.\ref{LF_parameters}, where $\phi$ is the LF amplitude (in cosmic volume density), $M_*$ is the characteristic luminosity controlling exponential suppression at the bright end, and $\alpha$ is the faint end logarithmic slope. 

Having introduced how UV LF parameters could be determined at given averaged redshifts, we next consider redshift variation in UV LFs. This is relevant for wide redshift intervals such as $3.5\leq z\leq 5.5$, as redshift variation in UV LF could affect the accuracy of magnification bias predicted. In order to perform magnification bias over arbitrarily narrow redshift interval, we investigate how LF parameters vary with redshift by breaking full redshift range of interest - $1.2 \leq z \leq 5.5$ - into smaller redshift bins. Inspired by the fact a galaxy at redshift $z=2$ could be both assigned to $1.5-2.5$ sample or $1-2$ sample for statistical measurements, we allow here different smaller redshift bins to overlap. This way we loose 'statistical independence' between adjacent bins, but gain more LF sampling points and also enough number of detections in all bins. This in turn allow us to more robustly determine the UV LF-redshift relation, and it was verified that different binning choice lead only to a controlled variation ($\sim\pm3\%$) in the final contamination level we report. The redshift bin sizes chosen are $0.6$ for center redshift $z<2$, $0.8$ if center redshift is within $2\leq z_c<3$ and $1.0$ for center redshift $z_c \geq 3$.

In each bin, we separately determined the Schechter parameters which are presented in Table.\ref{LF_parameters}. Our measured parameters are found well parameterized by the following relation with redshift :

\begin{equation}
    \phi(z) = 10^{(-0.237\pm 0.050) z + (1.042 \pm 0.115)} \text{  [1e-3/mag/Mpc}^3],
\end{equation}

\begin{equation}
    M_*(z) = \begin{cases}
         (-1.237 \pm 0.230) z + (-17.772 \pm 0.415) \\
         \text{ if } z \leq 2.47\\
         (-0.052 \pm 0.139)(z-2.48) + (-20.798 \pm 0.154) \\ 
         \text{ if } z > 2.47
        \end{cases}
\end{equation},
\begin{equation}
    \alpha(z) =  (-0.127 \pm 0.045) z + (-1.237 \pm 0.103).
\end{equation}

\noindent Following this parametrization, in Fig.\ref{LF_zs} we plot UV LFs at all redshifts over $1.2\leq z\leq 5.5$ in their integrated form (to cope with Fig.\ref{low_LF}), with redshift indicated by the color bar. We also present the "trajectory" of $M_*$ with redshift as a black dashed line. Overall, we can see at higher redshifts the faint end slope is steeper, UV LF amplitude is lower, and that exponential suppression occurs at a brighter magnitude. These observed trends agree with existing UV LF measurements from literature, and we comment that our determined LF-redshift relations are akin to what \cite{Bouwens2021} obtained.

% at all redshift over $1.2\leq z\leq 5.5$ thus determined are plotted in Fig.\ref{LF_zs}, where we plotted the integrated UV LF as a function of limiting absolute magnitude thresholds at different redshifts as indicated by the color bar. We also present the 'trajectory' of $M_*$ with redshift as a black dashed line.  %But we note \cite{Bouwens2021} only used galaxies satisfying Lyman Break Galaxies-like criteria, whereas we have not performed any restriction on the color hence our UV LFs may better incorporate all types of galaxies. For this reason, we chose not to use the better statistic measurements from \cite{Bouwens2021} and use our derived LF-redshift relations for magnification bias analysis. %For completeness and comparison, however, we will also provide in Sec.\ref{Posbias_Sec} a contamination level estimated based on \cite{Bouwens2021} UV LFs. 

\begin{deluxetable}{cccc}[h]
\tabletypesize{\scriptsize}
\tablewidth{5pt} 
%\tablenum{1}
\tablecaption{Fitted Schechter parameters over various redshift intervals  \label{LF_parameters}}
\tablehead{
\colhead{z range} & \colhead{$\phi$} & \colhead{$\alpha$} & \colhead{$M_*$} \\ \colhead{} & \colhead{(1e-3/mag/Mpc$^3$)} & \colhead{} & \colhead{(AB mag)}
} 
%\colnumbers
\startdata 
1.20-2.40 & 4.715$\pm$0.388 & -1.374$\pm$0.042 & -19.909 $\pm$0.052 \\ 
3.50-5.50 & 0.459$\pm$0.198 & -1.886$\pm$0.142 & -21.468 $\pm$0.232 \\ 
\hline
1.20-1.80 & 6.322 $\pm$ 0.427 & -1.365 $\pm$ 0.032 & -19.542 $\pm$ 0.044 \\
1.45-2.05 & 3.517 $\pm$ 0.246 & -1.491 $\pm$ 0.027 & -20.081 $\pm$ 0.044 \\
1.70-2.30 & 3.687 $\pm$ 0.392 & -1.447 $\pm$ 0.046 & -20.153 $\pm$ 0.066 \\
1.85-2.65 & 2.337 $\pm$ 0.433 & -1.584 $\pm$ 0.072 & -20.433 $\pm$ 0.109 \\
2.10-2.90 & 1.281 $\pm$ 0.261 & -1.759 $\pm$ 0.060 & -20.870 $\pm$ 0.119 \\
2.35-3.15 & 2.228 $\pm$ 0.387 & -1.562 $\pm$ 0.070 & -20.692 $\pm$ 0.103 \\
2.50-3.50 & 1.754 $\pm$ 0.332 & -1.571 $\pm$ 0.076 & -20.898 $\pm$ 0.112 \\
2.75-3.75 & 2.693 $\pm$ 0.434 & -1.319 $\pm$ 0.095 & -20.621 $\pm$ 0.103 \\
3.00-4.00 & 1.717 $\pm$ 0.459 & -1.439 $\pm$ 0.136 & -20.924 $\pm$ 0.167 \\
3.25-4.25 & 0.209 $\pm$ 0.082 & -2.031 $\pm$ 0.070 & -22.048 $\pm$ 0.230 \\
3.50-4.50 & 0.870 $\pm$ 0.252 & -1.772 $\pm$ 0.105 & -21.096 $\pm$ 0.161 \\
3.75-4.75 & 1.353 $\pm$ 0.611 & -1.670 $\pm$ 0.198 & -20.814 $\pm$ 0.260 \\
4.00-5.00 & 2.186 $\pm$ 0.616 & -1.349 $\pm$ 0.197 & -20.501 $\pm$ 0.188 \\
4.25-5.25 & 1.780 $\pm$ 0.602 & -1.256 $\pm$ 0.254 & -20.586 $\pm$ 0.239 \\
4.50-5.50 & 0.832 $\pm$ 0.299 & -1.601 $\pm$ 0.156 & -21.019 $\pm$ 0.216 
\enddata
\end{deluxetable}
%These CLF-redshift relations were used for testing contamination in the Shipley 18+ catalogs, but they may not be applicable to other catalogs. That is CLF-redshift relations best be re-derived when using a different sample. 

% Surprisingly our obtained CLF-redshift relation is similar to what \cite{Bouwens2021} determined, despite with larger uncertainties as a result of much smaller sample size. In paricular they have used Lyman Break Galaxy like criteria and we have not. 

\subsection{Implementing Magnification bias}
\label{sec_mag_bias_equation}

%As previously discussed, magnification bias expresses two competing effects that gravitational lensing makes on the observed surface number density of galaxies (per unit solid angle of sky). First, lensing makes detectable faint galaxies by raising their brightness to above the detection threshold; or, equivalently, by effectively lowering the detection threshold. The anticipated increase in the observed surface number density of galaxies, however, is countered by the decrease in the physical area of sky probed as the latter also is magnified by lensing. Depending on how steeply the LF continues to fainter magnitudes, lensing magnification may result in a net increased (magnification bias termed as positive), decreased (negative magnification bias), or leaving no effect on the overall surface number density of galaxies in cluster fields relative to in parallel fields. 

%Previously, we denoted the observed surface number density of galaxies at redshift $z$ as a function of limiting magnitudes as $\Gamma(<m_{lim},z)$. For a lensing magnification factor of $\mu_z$, magnification bias states the lensed densities is related to the unlensed one as 

At a given redshift $z$ and a local magnification factor $\mu_z$, magnification bias predict the lensed surface number density of galaxies ($n_{len}$) to be 
\begin{equation}
n_{len} (z,\mu_z,m_{lim}) = \frac{\Gamma(<(m_{lim}+2.5\log_{10}\mu_z),z)}{\mu_z}.
\label{bias_equation}   
\end{equation}
\noindent where $\Gamma$ is the underlying luminosity function (in integral form) as measured from the parallel fields in Sec.\ref{sec_counts_construct}, so that $\mu_z \rightarrow 1$ corresponds to the unlensed limit (for which $n_{len}\rightarrow n_{o}$). As could be seen, gravitational lensing lowers the unlensed detection threshold $m_{lim}$ by $2.5\log_{10}\mu_z$, whereas the factor of $\mu_z$ in the denominator reminds us that the sky area is simultaneously reduced. Note depending on how fast $\Gamma(<(m_{lim}+2.5\log_{10}\mu_z))$ increases as $\mu_z$ increases, lensing magnification may result in a net increase (with $n_{obs}> n_o$, corresponding to positive magnification bias), decrease ($n_{obs}< n_o$, negative bias), or leaving no change on the observed surface number density of galaxies ($n_{obs} = n_o$, critical bias).

In the remainder of this paper, we will predict for the surface number density of galaxies over $3.5\leq z\leq 5.5$ at different levels of lensing magnifications. We then test for contaminants as any excess population observed in the cluster fields. To measure the observed surface number density at different levels of lensing magnification, we divide cluster field FOVs into ten magnification bins according to the following lens models: (1) v4 CATS \textit{lenstool} models \citep{Caminha2016, Caminha2017, Jauzac2016, Lagattuta2017, Limousin2016, Mahler2018}, adopted as a representative of parametric lens modeling techniques; (2) v4.1 WSLAP+ models \citep{Lam2014, Diego2015a, Diego2015b, Diego2016a, Diego2016b, Diego2018}, representative of free-form lens modeling techniques; and (3) additional two state-of-the-art parametric glafic models\footnote{These glafic models are not to be confused with existing ones on HFF archive. Our adopted models are with most updated lensing constraints, and are carefully examined for their predictive power.} available only for MACS0416 (provided by collaborator) and Abell370 \citep{Keith2022}. To ensure comparable statistics, magnification bins are equally spaced in logarithmic scale with a logarithmic width of 0.2 up to a magnification factor of 100.\footnote{Galaxies with predicted magnification factor above 100 have large uncertainties owing to sitting very close to critical curves, such highly magnified sources are all put into the highest magnification bin available.} 

\begin{figure}
\includegraphics[width=0.46\textwidth]{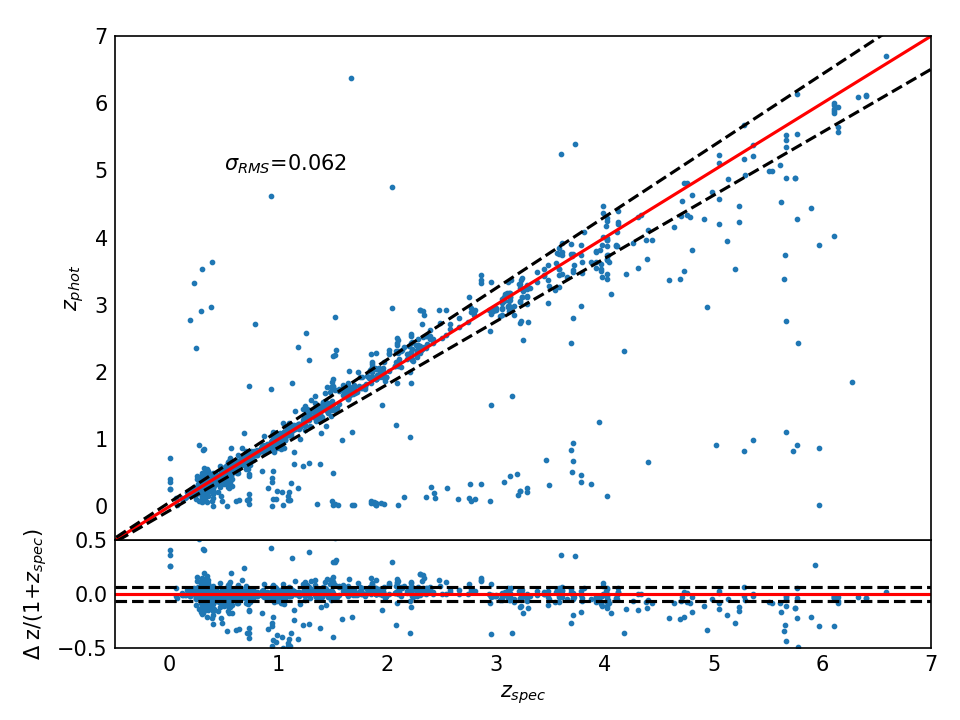}
\caption{Spec-z plotted against reported photo-z for galaxies from S18 catalogs (combining all HFF cluster fields) to determine $z_{RMS}$. The normalized deviations $(z_{phot}-z_{spec})/(1+z_{spec})$ are provided in the bottom panel, and the scatter range is seen well captured by the black dash lines (corresponding to $\pm 0.062$ and $z_{RMS} \sim 0.062(1+z)$).}
\label{specz_vs_photz_shipley}
\end{figure}

At any particular redshift $z$, we predict the surface number density of galaxies in each magnification bin by collecting the number of galaxies predicted on each image plane pixel, and subsequently divide this number by the total area. Similarly, for the same magnification bin, we measure the observed surface number density of galaxies by counting the number of UV bright galaxies (see Sec.\ref{subsec_uv_bright_sample}) over redshifts $z \pm z_{RMS}$. Here, $z_{RMS}$ is the typical scatter (reflective of uncertainties) of reported photo-z about spec-z for S18 catalogs, and we measured it to be $z_{RMS} \sim 0.062(1+z)$ from Fig.\ref{specz_vs_photz_shipley}. Correspondingly, when calculating both the observed and predicted surface number density over a wide redshift range (e.g. $1.2\leq z\leq 2.4$ and $3.5\leq z\leq 5.5$), we use redshifts separated by $z_{RMS}$ to individually implement magnification bias. Obtained surface number densities at different redshifts are later added up to yield full density for the wide redshift range. 

With the same philosophy and same set of magnification bias maps at different redshifts, we can also predict surface number density in different radial bins. This is how the shown prediction in Fig.\ref{cluster_member_distribution} were obtained.

% For all the pixels corresponding to each magnification bin, we add up the predicted number of galaxies contained within these pixels and divide this number by the total area of thse pixels to obtain the predicted surface number density of galaxies (at a given redshift z).To calculate the observed galaxy density from catalogs in the same magnification bin, we count all galaxies having redshifts within $z \pm z_{RMS}$, where $z_{RMS}$ reflects the typical uncertainty of the photo-z at redshift $z$. We determine $z_{RMS}$ by plotting estimated photo-z of galaxies in S18 catalogs against their cross-matched spec-z value (see \cite{Shipley2018} for more details) in Fig.\ref{specz_vs_photz_shipley} (combining all available HFF cluster fields). The normalized deviations $(z_{phot}-z_{spec})/(1+z_{spec})$ are shown in the bottom panel, and its scatter range is seen well captured by the black dash lines corresponding to $\pm 0.062$, giving $z_{RMS} \sim 0.062(1+z)$. 

\subsection{UV Bright Sample}
\label{subsec_uv_bright_sample}

\begin{figure*}
\includegraphics[width=0.9\textwidth]{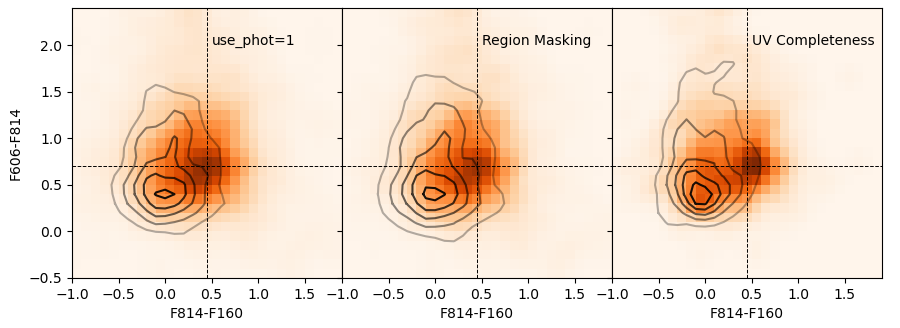}
\caption{Color-color after different stage of selection.}
\label{color_check_selection_bias}
\end{figure*}

%Within S18 catalogs, we only use galaxies flagged with \texttt{use$\_$phot}=1, i.e. these are non-star-like galaxies considered to have reliable photometries and photo-z estimations. 

In Table.\ref{sample_remaining}, the number of \texttt{use$\_$phot}=1 galaxies from S18 catalogs over redshift ranges $1.2\leq z\leq 2.4$ and $3.5\leq z\leq 5.5$ are listed under \textit{use$\_$phot=1} column. The sub-sample after applying exclusion regions are listed under the \textit{Region Masking} column, it is seen this selection leaves about $30\%$ of sample in the cluster fields and $40\%$ in the parallel fields. We demonstrate in Fig.\ref{color_check_selection_bias} that this sub-sample is reflective of the full \texttt{use$\_$phot}=1 sample. Here background histograms and black contours are F606-F814 vs F814-F160 distribution of $3.5\leq z\leq 5.5$ galaxies in the cluster and parallel fields respectively. We also marked the peak of cluster field distribution by the dashed cross, which is not seen to vary as we restrict to \textit{Region Masking} sample. 

As magnification bias analysis relies on UV luminosity functions, we in fact only need UV bright galaxies in the band covering their rest frame UV wavelengths. As discussed in Sec.\ref{sec_counts_construct}, this rest frame wavelength is chosen to be $1600\AA$ at $z\leq2$ but $2000\AA$ at $z<2$. The final analysis sample thus comprise of galaxies brighter than data completeness thresholds in the respective band. The corresponding sample sizes are listed in the \textit{Completeness} column of Table.\ref{sample_remaining} for both redshift ranges. Again, in the last panel of Fig.\ref{color_check_selection_bias} we demonstrate that restricting to this sub-sample does not alter significantly the color-color distribution. I.e. overall we are left with $\sim 17-18\%$ of \texttt{use$\_$phot}=1 galaxies for analysis. But as we showed in Fig.\ref{color_check_selection_bias}, the UV bright sample is representative of the whole \texttt{use$\_$phot}=1 sample. In another words, the contamination inferred using only the UV bright sample is expected to be reflective of contamination level for the full galaxy sample.

%In Table.\ref{data_complete_thresholds_shipley}, we indicate in the second row over what redshift range each band are used for data selection, and the number of sources remained after selection are listed in the \textit{Completeness} column of Table.\ref{sample_remaining}. 

\begin{deluxetable}{ccccccc}
\tabletypesize{\scriptsize}
\tablewidth{5pt} 
%\tablenum{1}
\tablecaption{Number of $1.2\leq z\leq 2.4$ and $3.5\leq z\leq 5.5$ sources left from S18 catalogs after each selection step. \label{sample_remaining}}
\tablehead{
\colhead{Field} & \colhead{Redshift} & \colhead{\texttt{use$\_$phot}=1} & \colhead{Region Masking} & \colhead{Completeness}}
%\colnumbers
\startdata 
A2744clu & 1.2-2.4 & 1593 & 491 & 238 \\
 & 3.5-5.5 & 917 & 382 & 148 \\
A2744par & 1.2-2.4 & 1751 & 765 & 326 \\
 & 3.5-5.5 & 454 & 211 & 69 \\
M0416clu & 1.2-2.4 & 1822 & 720 & 319 \\
 & 3.5-5.5 & 839 & 507 & 120 \\
M0416par & 1.2-2.4 & 2224 & 925 & 344 \\
 & 3.5-5.5 & 506 & 317 & 75 \\
M0717clu & 1.2-2.4 & 966 & 361 & 159 \\
 & 3.5-5.5 & 582 & 261 & 140 \\
M0717par & 1.2-2.4 & 1593 & 745 & 330 \\
 & 3.5-5.5 & 329 & 197 & 80 \\
M1149clu & 1.2-2.4 & 1561 & 539 & 266 \\
 & 3.5-5.5 & 680 & 320 & 101 \\
M1149par & 1.2-2.4 & 1761 & 847 & 376 \\
 & 3.5-5.5 & 392 & 228 & 79 \\
AS1063clu & 1.2-2.4 & 1614 & 482 & 201 \\
 & 3.5-5.5 & 673 & 259 & 87 \\
AS1063par & 1.2-2.4 & 1895 & 952 & 427 \\
 & 3.5-5.5 & 398 & 191 & 70 \\
A370clu & 1.2-2.4 & 1222 & 395 & 165 \\
 & 3.5-5.5 & 873 & 358 & 169 \\
A370par & 1.2-2.4 & 1682 & 850 & 381 \\
 & 3.5-5.5 & 409 & 224 & 66 \\
 \hline
Cluster total & 1.2-2.4 & 8778 & 2988 & 1348 \\
 & 3.5-5.5 & 4564 & 2087 & 765 \\
Parallel total & 1.2-2.4 & 10906 & 5084 & 2184 \\
 & 3.5-5.5 & 2488 & 1368 & 439 \\
\enddata
%\tablecomments{}
\end{deluxetable}

\begin{figure*}
    \includegraphics[width=0.97\textwidth]{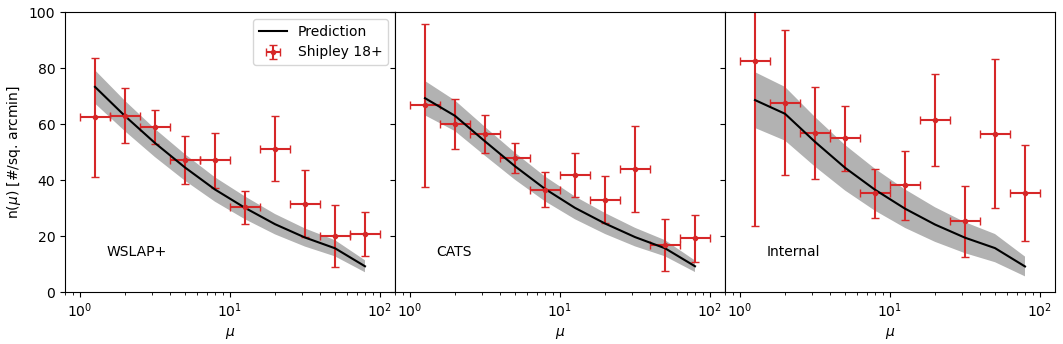}
    \caption{Magnification bias analysis over $1.2<z<2.4$ for S18 catalogs. Predicted magnification bias (black line) as function of magnification factor $\mu$ compared with derived densities (red data points) when adopting glafic, WSLAP+ and cATS models respectively. Data points are after averaging over available cluster fields for each catalogs. The shaded region for prediction reflects measurement uncertainty in CLF-redshift relations. Predicted surface density is seen to decrease as magnification increases, indicating a negative magnification bias. The derived densities is seen to follow a similar trend as prediction, but may with a systematic offset. We also notice data in some magnification bins are severely deviating from the predicted level, suggesting we have contamination in those bins.}
    \label{neg_bias_magbin_diego_collective}
\end{figure*}

\begin{figure*}
     \centering
     \includegraphics[width=0.97\textwidth]{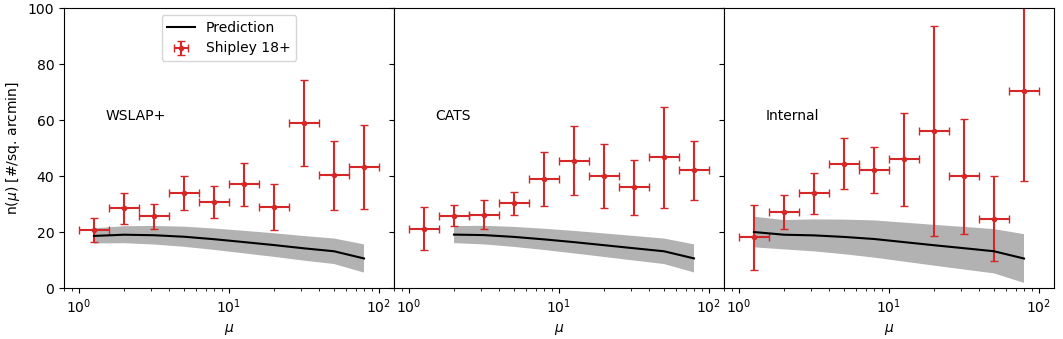}
     \caption{Magnification bias analysis over $3.5\leq z \leq 5.5$ for S18 catalogs. Predicted surface density is seen to decrease as magnification increases, indicating a negative magnification bias. On the other hands, the cataloged sources are seen to be more populated in the higher magnification regions, demonstrating strong contamination. }
        \label{pos_bias_magbin_diego_collective}
\end{figure*}

\section{Predicted vs Observed galaxy density}
\label{bias}

We start by testing our methodology over the redshift range of $1.2<z<2.4$, which we expect to suffer no contamination by early-type cluster members (Sec.\ref{sec_counts_construct}). Fig.\ref{neg_bias_magbin_diego_collective} plots the predicted surface number densities in different magnification bins as black solid curves for each lens model. The densities are averaged over available HFF cluster fields to mitigate cosmic variance effects. For the WSLAP+ and CATS lens models, we had all HFF fields available, but for the Internal models, only the M0416 and A370 fields were used. This results in larger uncertainties (shaded region) for the Internal model predictions. As a reminder, the predicted densities are based on UV luminosity function redshift relations derived from parallel fields. The shaded regions reflect the uncertainty in the luminosity function propagated into the magnification bias estimates. While the different lens models predict slightly varying density distributions, they all decrease toward higher magnification regions, indicating an negative magnification bias.

We compare our predictions to the observed surface number densities from the UV bright sample, which are plotted as red data points in Fig.\ref{neg_bias_magbin_diego_collective}. The measured densities also average over available cluster fields, with vertical error bars incorporating Poisson statistics and variance between fields. The observed densities show the same trend of decreasing toward higher magnification regions, consistent with the predicted negative magnification bias. Referring back to Fig.\ref{excess}, negative bias thus provides a natural explanation for the relative deficit of galaxies in the cluster fields observed around $z\sim2$. 

While the predicted and observed trends agree regardless of lens model, a strong excess is observed in high magnification regions $\mu >10$. This may indicate contamination. For instance, these could be faint cluster members with poorly-estimated redshifts, and are naturally more populated towards inner higher magnification regions.

Fig.\ref{pos_bias_magbin_diego_collective} shows a similar magnification bias analysis over redshift range $3.5\leq z\leq 5.5$. In contrast to the strong negative bias shown in Fig.\ref{neg_bias_magbin_diego_collective}, the predicted magnification bias (black curve) is only mildly negative. The measured galaxy surface number densities (red points), on the other hand, display a strong increasing trend towards higher magnification region. This deviation supports our previous hypothesis (Sec.\ref{section_implications}) that contaminants may be misidentified cluster members. 

The excess above predictions provides us a measure of contamination level. We calculate this by collectively counting the total measured and predicted number of galaxies over $3.5\leq z \leq 5.5$. This way, the contamination levels are estimated to be $59.38 \pm 11.12 \%$ based on CATS models, $53.48 \pm 10.33 \%$ based on WSLAP+ models, and $59.53 \pm 18.52 \%$ based on Internal models. We note that regardless of lens model, more than half of $3.5\leq z \leq 5.5$ galaxies in the S18 catalogs would be contaminants from misidentified cluster members! Furthermore, we remind readers that the inner regions of clusters may have a non-uniform, shallower detection threshold (see Sec.\ref{sec_data_complete}). This would instead imply the galaxy sample in higher magnification bins is incomplete, hence the number of lensed galaxies is over-predicted. As such, the actual contamination levels may also be higher than what are reported above. 

While the precise contamination level is hard to obtain, we have demonstrated severe contamination ($>50\%$) in the S18 UV bright sample over $3.5\leq z\leq 5.5$. As the UV bright sample reflects the full \texttt{use\_phot}=1 sample (see Sec.\ref{subsec_uv_bright_sample}), we expect also a similar contamination level for entire S18 catalogs.

\section{Discussion and Conclusion}
\label{conclusion_section}

In this article, we demonstrated that photo-z catalogs suffer from severe contamination, most likely by misidentified cluster members. We started by proving the existence of such a contamination via color diagnostics, followed by the first quantitative estimation made possible by magnification bias. To our great surprise and worry, the estimated contamination level the redshift range of $3.5\leq z\leq 5.5$ is as high as $\sim 60\%$! And on average (over estimates based on different lens model, using number of cluster fields available as weights), the contamination level is found to be $56.87 \pm 11.84 \%$. Such a severe contamination level, as we demonstrate in Sec.\ref{contaminated_LF_sec}, may result in non-physical turn-ups in the cluster field UV LFs if not properly mitigated. As a consequence, contaminants could hinder the test of faint end turnovers in UV LFs, such as those predicted by baryonic feedback models, or by alternative models for dark matter (e.g. warm or wave dark matter). The prospects of mitigating contaminants are discussed in Sec.\ref{sec_mitigating}. We start by investigating the effectiveness of commonly adopted Lyman Break Galaxy-like selection criterion, and end our discussion by stating the best hope for individual mitigation is by incorporating deep JWST measurements.

\subsection{Potential up-turns in LFs?}
\label{contaminated_LF_sec}

\begin{figure*}
\gridline{\fig{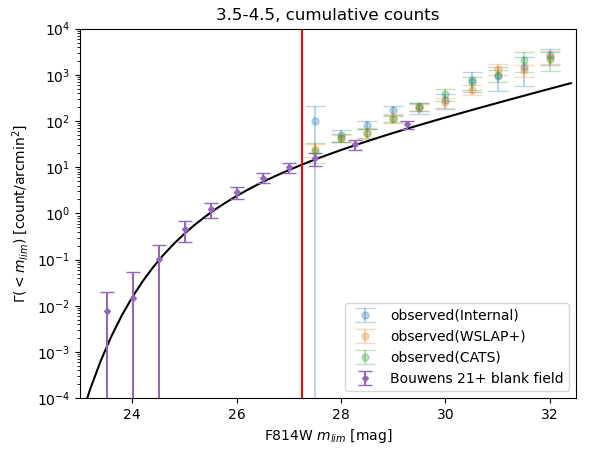}{0.45\textwidth}{(a)}
\fig{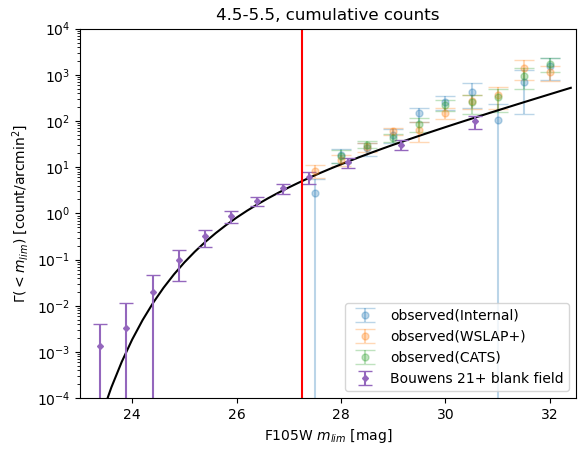}{0.45\textwidth}{(b)}
}%\fig{figs_pos/clus_field_data_vs_para_cumu_counts_1.2_1.8.png}{0.3\textwidth}{(c)}
\caption{Cumulative UV LF implied by the cluster field densities compared with deeper parallel field measurements from \cite{Bouwens2021} over redshifts (a) 3.5-4.5 and (b) 4.5-5.5. The implied values are obtained by multiplying measured surface number density in each magnification bin with respective mean magnification factor, and are shown as colored data points depending on lens model involved. Owing to lensing magnification, these implied values are all fainter than the data completeness threshold (red vertical line) in the band capturing rest frame 1600$\AA$. We plot also our derived LF (integrated and extrapolated to fainter magnitudes) at median redshift 4 and 5 from parallel fields as black solid line for comparison. As could be seen, implied values rise above derived LF immediately after the data completeness limit, suggesting a need for faint end up-turns in UV LFs. But having such up-turns is seen not consistent with the deeper blank field measurements from \cite{Bouwens2021}, which are shown as purple data points. }
\label{reproducing_data}
\end{figure*}

Recall from Fig.\ref{cm_high_z}, the concentration (dominated by contaminants) to cluster red sequence occurs mainly at faint magnitudes. In terms of UV LF construction, this may lead to heavily over-estimated values at the faint end if contaminants are not properly dealt with. Such an impact is demonstrated by Fig.\ref{reproducing_data}, in which we reverse engineered\footnote{Multiplying measured surface number density of galaxies with mean magnification factor gives LF integrated up to the respective lowered threshold.} magnification bias equation Equ.\ref{bias_equation} to obtain implied cumulative LF values (colored data points) fainter than the unlensed data completeness threshold (red vertical line). As could be seen, the implied value is significantly higher than both (1) faint end extrapolation of our derived UV LF from parallel fields, shown as black solid lines; and (2) more recent blank field measurements\footnote{What \cite{Bouwens2021} measured was LFs. Cumulative LFs were calculated from LF measurements using the Simpson's rule.} from \cite{Bouwens2021}, shown as purple data points, over the same redshift range and covering magnitudes fainter than S18's data completeness threshold. This (having contamination) may explain why some UV LF measurements from cluster fields (e.g. \cite{Livermore2017,Bouwens202205}) are elevated at the very faint end. Another implication, in terms of cosmological model testing, would be contaminants may wash out faint end turnovers in UV LFs and hinder the test of cosmological models. 

\subsection{Prospect of Mitigation}
\label{sec_mitigating}

We recall in literature, UV LF are often constructed with the aid of Lyman Break Galaxy-like (LBG-like) selection criteria which is color-based and independent of photo-z measurements. To investigate whether such selection rules could statistically mitigate contaminants in the lensing fields, we repeat the same magnification bias analysis over $3.5\leq z\leq 5.5$ using only LBG-like galaxies selected by criteria\footnote{We apply the corresponding LBG-like criteria separately to $3.5\leq z \leq 4.5$ and $4.5 \leq z \leq 5.5$ samples, and to both cluster fields and parallel fields. } reported by \cite{Bouwens202203}. 

These LBG-like criteria were found very stringent and left us with only $\sim 34\%$ of the UV bright sample, rendering it hard to re-measure the UV LF from the parallel fields. We thus adopt the blank field UV LFs (\cite{Bouwens202203} is over lensing fields) derived by \cite{Bouwens2021} which were also based on LBG-like galaxies. For the LBG-like criteria selected sample, the new (average) contamination level is found to be $11.04 \pm 11.79\%$. This is much lower than previously reported and compatible with zero, signaling a potential success. We warn readers, however, while the selected sample is arguably 'cleaner' it is also less complete. In particular, LBG-like condition select dominantly blue galaxies, hence UV LF derived based on LBG-like galaxies might not be optimal for tests involving intrinsically faint/red high-z galaxies.

%\footnote{Technically, \cite{Bouwens2021} had slightly different LBG-like criteria to \cite{Bouwens202203}. But we have checked \cite{Bouwens2021} UV LFs to correctly describe the luminosity distribution of \cite{Bouwens202203} LBG-like galaxies. We also checked that \cite{Bouwens2021} UV LFs are compatible with our measurements from LBG-selected parallel field samples at the faint magnitudes.}

%In particular, from Fig.\ref{color_sB_con_on_LBG} we see galaxies that LBG-like galaxies in the cluster fields (whose distribution are indicated by blue contours) are those with dominantly blue color in F814-F160. Hence the high-z galaxies that are intrinsically red would be missed. As a consequence, UV LF derived based on LBG-like galaxies might not be optimal for tests involving intrinsically faint high-z galaxies (which could be intrinsically red). Moreover, the blue contours straddle a region in between the cluster field concentration with parallel fields, leading to worry whether some contaminants are still present. 

%\begin{figure}
%\includegraphics[width=0.46\textwidth]{figures_weed/csb_3.5-5.5_814160_NoDiff_LBG.png}
%\caption{LBG-selected galaxies on color-surface brightness plots. The distribution of LBG-selected galaxies are marked by blue contours. }
%\label{color_sB_con_on_LBG}
%\end{figure}

The best hope to individually mitigate interlopers thus lies in attaining deeper observations near rest frame $\sim 4000\AA$ of the high-z galaxies, such that their Balmer breaks could be observed. This could be achieved, for instance, with the aid of the newly launched JWST. Examples of such helpful NIRCam observations on HFF clusters include those from PEARLS (GTO 1176, \cite{Windhorst2022} targets include M0416, M1149, A2744), UNCOVER (GO 2561, \cite{2022arXiv221204026B}, targets A2744) and CANUCS (GTO 1208, \cite{2022PASP134b5002W}, targets include M0416, M1149) team. These NIRCam observations could further be complemented with spectroscopic observations on these clusters such as those by GLASS (ERS 1324, \cite{Treu2022}) and CANUCS team for better lens model construction and a more robust mitigation of the contaminants. With these observations, we could also address the contamination level in nearby blank fields (if covered by both JWST and HST) which we have not addressed in this paper.

%Nevertheless, misidentification problems could still occur for the newly detected faint sources in JWST without HST measurements. In particular, dusty lower redshift galaxies at $z\sim5$ could be mistaken to be galaxies at much higher redshift at $z\sim16$ as have been proved by \cite{2023arXiv230315431A}. Worse, this problem is no longer restricted to cluster fields as parallel fields could be equally severe. Under such circumstances, magnification bias would still be useful for cross-checking the number of extreme high-z galaxies detected with what cosmological models predict. That is, as soon as reliable lens models become available. 

\section{acknowledgement}

The authors thank Arsen Levitskiy and Tim Carleton for many useful and insightful discussions. J.Z. acknowledge the financial support by Presidential Scholarship from the University of Hong Kong. J.L., T.B., G.M. and S.K.L. acknowledge the Collaborative Research Fund under grant C6017-20G, which is issued by the Research Grants Council of Hong Kong SAR. J.Z., J.L. and S.K.L acknowledge also RGC/GRF 17312122 issued by the Research Grants Council of Hong Kong SAR. R.A.W. acknowledges support from NASA JWST Interdisciplinary Scientist grants NAG5-12460, NNX14AN10G and 80NSSC18K0200 from GSFC.

\appendix

\bibliography{reference}{}
\bibliographystyle{aasjournal}

\end{document}